\documentclass[12pt]{article}
\usepackage{axodraw}

\newcommand{\nn}{\nonumber}

\newcommand{\cF}{{\cal F}}

\def\bc{\begin{center}}
\def\ec{\end{center}}
\def\bi{\begin{itemize}}
\def\ei{\end{itemize}}
\def\be{\begin{equation}}
\def\ee{\end{equation}}
\def\bea{\begin{eqnarray}}
\def\eea{\end{eqnarray}}

\def\bdm{\begin{displaymath}}
\def\edm{\end{displaymath}}
\def\ds{\displaystyle}

\begin{document}

\author{ Rogalyov R.N.\footnote{e-mail: rnr@ihep.ru }\\
        {\it Institute for High Energy Physics}\\
         {\it Protvino, Moscow region, 142281, Russia}}
\title{Two-Loop Fermionic Integrals
in Perturbation Theory on a Lattice  \\
         }

\maketitle


\begin{abstract}
A comprehensive number of one-loop integrals in a theory with
Wilson fermions at $r=1$ is computed using the Burgio--Caracciolo--Pelissetto algorithm.
With the use of these results, the fermionic propagator in the
coordinate representation is evaluated, thus making it possible
to extend the L{\" u}scher-Weisz procedure for two-loop integrals to the
fermionic case. Computations are performed with FORM and REDUCE packages.
\end{abstract}

\vspace*{11mm}

\bc 
{\small Talk at the 13th International Workshop\\
 on Advanced Computing and Analysis Techniques in Physics Research, \\
		February 22-27, 2010, \\
			Jaipur, India}
\ec

\thispagestyle{empty}

\newpage

\section{Introduction}

Perturbative calculations in lattice gauge theories (for a review, see \cite{CapitaniPR}) are of interest from several points of view.

Firstly, they are needed to determine the $\Lambda_{LAT}$ parameter of QCD in the 
lattice regularization and its relation to the respective value $\Lambda_{QCD}$ in the continuum theory. 

Secondly,  every lattice action defines a specific regularization scheme, 
and thus one needs a complete set of renormalization computations 
in order for the results obtained in Monte Carlo simulations 
be understood properly. Perturbation theory is required to 
establish the connection of the matrix elements computed on a lattice 
with their values in the continuum theory \cite{Braun}, \cite{Hagler}. 
In this connection, it should be emphasized that the use of one-loop perturbative renormalization constants gives rise 
to large systematic uncertainties in lattice calculations of the momenta of hadronic structure functions 
\cite{Hagler} and respective two-loop computations are needed.

Thirdly, perturbative calculations provide the only 
possibility for an analytical control over the continuum limit in QCD.
One can also mention anomalies, proof of renormalizability, 
Symanzik improvement program and other fields of application
of lattice perturbation theory.

Here we consider one- and two-loop diagrams with Wilson ($r=1$) fermions
at zero external momenta \cite{Kawai}.
We outline the Burgio-Caracciolo-Pelissetto (BCP) method \cite{Burgio} of calculations of 
one-loop integrals and describe the respective computer algorithm \cite{Rogalyov}.
This algorithm allows to compute the fermionic propagator in the coordinate
representation and, therefore, to extend the L{\" u}scher-Weisz (LW) method \cite{Luscher}
to the fermionic case; such extension is presented in Section~4.

\subsection{Notation}
We use the following designations: $\tilde n$ stands for the set $n_1,n_2,n_3,n_4$;
$\ x=(x_1,x_2,x_3,x_4)$, where $x_\mu$ are 
integer-valued coordinates of an infinite four-dimensional lattice 
$\Lambda=\{x:\ x_\mu\in Z\!\!\!Z\}$; we also need the lattice $\Lambda' = \Lambda \backslash \{0\}$ with removed site x=(0,0,0,0);  
\be 
|x| = |x_1| + |x_2| + |x_3| + |x_4|,  \qquad  \qquad  \qquad [x^n] = x_1^n +x_2^n +x_3^n +x_4^n.
\ee
Then we give the expressions for the denominators of bosonic and fermionic propagators,
\bea
\Delta_B(k) &=& 4+\mu_R^2-\cos(k_1 )-\cos(k_2 )-\cos(k_3 )-\cos(k_4); \\ \nonumber
\Delta_F(k) &=& 10-4\;\sum_{\mu=1}^4 \cos(k_\mu) + \sum_{1\leq \mu < \nu \leq 4} \cos(k_\mu) \cos(k_\nu) + \mu_R^2 \nonumber
\eea
where $\mu_R$ is the fictitious mass for infrared regularization.
We also use $D_F=2\;\Delta_F$ and $D_B=2\;\Delta_B $ normalized in the standard way ($D_{B(F)(k)}\simeq 1/k^2$ as $k\to \infty$).
These propagators in the coordinate representation are defined as follows:
\be 
 G_{B(F)}(x) =  \int_{BZ} {dk \over (2\pi^4)} {e^{-ikx} \over D_{B(F)}(k) },
\ee
where $BZ$ is the Brillouin zone, $\ds BZ = \Big\{ p:\ -\;{\pi\over a}\leq p_\mu \leq {\pi\over a} \Big\}$;
\be
\hat p_\mu = {2\over a} \sin \left({ p_\mu a \over q }\right), \qquad \qquad \hat p^2 = \sum_{mu=1}^4 \hat p_\mu^2;
\ee
$a$ is the lattice size. In this work we set $a=1$ for the sake of simplicity.

\section{The Burgio--Caracciolo--Pelissetto method}
\subsection{Bosonic Intefgrals}
\noindent The integrals under study are defined as follows: $F(q;\tilde n) = \lim_{\delta \to 0} F_\delta(q;\tilde n)$, where
\be\label{InitialBosonInt}
 F_\delta(q;\tilde n) 
=  \int_{BZ} dk\;{\cos(k_1 )^{n_1} \cos(k_2 )^{n_2} \cos(k_3 )^{n_3} \cos(k_4 )^{n_4} \over \Delta_B^{q + \delta}}.
\ee
Here  $\delta$ is an infinitesimal parameter for an intermediate regularization \cite{Burgio}.
This parameter makes it possible to derive\footnote{Using integration by parts} the recursion relations of the form
\bea\label{RRforFbosonic}
& & F(q;..., n_\mu,...) = F(q;..., n_\mu-2,...) \;-\; \\ \nonumber
 && - { (n_\mu-1) F(q-1;...,n_\mu-1,...)  \over q-1+\delta} + { (n_\mu -2) F(q-1;...,n_\mu-3,...) \over q-1+\delta} \qquad (n_\mu \geq 2). \nonumber
\eea
With these relations and similar relations for $n_\mu \leq 1$, one can
express the integrals (\ref{InitialBosonInt}) in terms of the quantities
\be
G_\delta (q,\mu_R^2)=\int_{BZ} {dk \over (2\pi)^4 } {1\over (\Delta_B)^{q+\delta}}.
\ee
Up to terms of the order ${\cal O}(\mu_R^2)$ and  ${\cal O}(\delta)$, this expression has the form
\be\label{BosFtoGgen}
F_\delta (q,\tilde n) = 
\sum_{r=q-n_1-n_2-n_3-n_4}^{0} A^{(-)}_{qr}(\delta,\tilde n) G_\delta (r,0) +
\sum_{r=1}^{q} A^{(+)}_{qr}(\mu_R^2,\tilde n) G_\delta(r,\mu_R^2), 
\ee
where $A^{(-)}_{qr}(\delta,\tilde n)$ have a pole singularity in  $\delta$, and
$A^{(+)}_{qr}(\mu_R^2,\tilde n)$ are polynomials in $\mu_R^2$.
As for the function $G_\delta(r,\mu_R^2)$, the domains $r>0$ and $r \leq 0$
should be considered separately.
At $r>0$, $\ \delta$ can be safely set to zero 
and the function $G_\delta(r,\mu_R^2)$ should be expanded in powers of $\mu_R^{-2}$:
\be \label{G_FMRdef1}
G_\delta (r,\mu_R^2) =  {1\over (2\pi)^2 \Gamma(r)} \left[ \; - \; b_{r-2} l_C + \sum_{k=1}^{r-2} { b_{r-k-2} \Gamma (k) \over (\mu_R^2)^k} \right] \; + \; J(r)\; + \; {\cal O}(\mu_R^2) + {\cal O}(\delta),
\ee
where $b_n$ are the coefficients of the asymptotic expansion at $z\to \infty$ of the function\footnote{$I_0(z)$ is the Infeld function.}
\be\label{InfeldAsExp0}
\exp(-4z) I_0^4(z) \simeq {1\over (2\pi z)^2}
\left(1+ {b_1\over z} + {b_2\over z^2} + ... \right),
\ee
$l_C = \ln (\mu_R^2) + C$, and $C$ is the Euler-Mascheroni constant.
At $r<0$, $\mu_R$ can be safely set to zero and the function $G_\delta(r,0)$ should be expanded in $\delta$ as follows: $G_\delta(r,0)= B(r) + J(r)\delta + {\cal O}(\delta^2)$.

The functions $J(q)$, in their turn, obey recursion relations of the type 
\be\label{RRboson}
c_0(q) J(q) + c_1(q) J(q+1) + c_2(q) J(q+2) + c_3(q) J(q+3) + c_4(q) J(q+4) = 0
\ee
derived in \cite{Burgio}; the explicit expressions for the coefficients $c_i(q)$
can be found in \cite{Rogalyov}. Thus we express $J(q)$ at $q\geq 4$ and at $q\leq 0$ 
in terms of $J(0), J(1), J(2)$ and $J(3)$. 
It should be noted that $J(0)$ does not appear in ultimate expressions for the integrals (\ref{InitialBosonInt}).
Then one can introduce the values
\be
Z_0 = {J(1)\over 2}, \qquad F_0 = 4\pi^2 J(2), \qquad Z_1 =  32 J(3) - 8 J(2) + {13 \over 6 \pi^2} + {1\over 4},
\ee
which are equal to \cite{CapitaniPR} $ Z_0\approx0.15493339023, Z_1\approx0.10778131354,  F_0\approx4.369225233874758$.

\subsection{Fermion Integrals}
In the fermionic case, we consider the quantities
$F(p,q;\tilde n) = \lim_{\delta \to 0} F_\delta(p,q;\tilde n)$, where 
\be 
F_\delta(p,q;\tilde n) = \lim_{\delta \to 0} \int {d^4k\over (2\pi)^4} 
{\cos^{n_1}(k_1) \cos^{n_2}(k_2) \cos^{n_3}(k_3) \cos^{n_4}(k_4) \over \Delta_B^q
\Delta_F^{p+\delta}}.
\ee
With the recursion relations similar to (\ref{RRforFbosonic}),
these integrals are expressed in terms of the functions $\ds G_\delta(p,q) = \int {d^4k\over (2\pi)^4} {1 \over \Delta_B^q \Delta_F^{p+\delta}},$
which can be represented in the form 
\bea\label{GdeltaExpansion}
G_\delta(p,q)&=& D(p,q;\mu_R^2) + B(p,q) + \delta \;(L(p,q;\mu_R^2)+J(p,q)) + O(\delta^2) , \qquad p\leq 0; \\ \nonumber
G_\delta(p,q)&=& D(p,q;\mu_R^2) + J(p,q) + O(\delta), \qquad p > 0.  \nonumber
\eea
The divergent parts $D(p,q;\mu_R^2)$ and $L(p,q;\mu_R^2)$ in the domain of interest can be determined by 
a straightforward procedure \cite{Burgio}, whereas the functions
$B(p,q)$ and $J(p,q)$ obey recursion relations of several types.
These relations and the procedure of their derivation were described in \cite{Burgio};
their explicit form (very cumbersome) is given in \cite{Rogalyov}.
With the use of these relations, the functions $F(p,q;\tilde n)$
can be represented (see \cite{CapitaniPR}, \cite{Burgio}) 
as linear combinations of the constants $F_0$, $Z_0$, $Z_1$ and
\bea\label{Ydef2}
Y_0 &=& {J(2,0)\over 4}\; - \; {F_0 \over 16\pi^2}, \qquad Y_1 = {1 \over 48} - {1 \over 4}\; Z_0 - {1 \over 24}\; J(-1,2) + 
{1 \over 12}\; J(0,1) + {1 \over 12}\; J(1,0), \\ \nonumber
Y_2 &=&  { 1 \over 6 } - {1\over \pi^2} - Z_0 - { 1 \over 6 }\; J(-1,2) + 
{1 \over 3}\; J(0,1) -{ 1 \over 24} \; J(1,-2) - {1 \over 12} \; J(1,-1) - \\ \nonumber
&& - 
{17 \over 8}\; J(1,0) + 4\; J(1,1) - {1 \over 48}\; J(2,-2) + {25 \over 6}\; J(2,-1)
 - 4\; J(2,0), \\ \nonumber
Y_3 &=&  - {1\over 384\pi^2} - F_0\; {1\over 128\pi^2} + {1 \over 96}\; Z_0 - 
{1 \over 48}\; J(-1,3) + {1 \over 192}  \; J(0,1) + {1 \over 48}\; J(0,2) + {1 \over 48}\; J(1,1);
\eea
\bea\label{Ydef1}
&& Y_4={J(1,0)\over 2},  \qquad Y_5 = J(1,-1), \qquad Y_6 = 2 J(1,-2), \qquad Y_7={J(2,-1)\over 2}, \\ \nonumber
&& Y_8 = J(2,-2),  \qquad Y_9 = {J(3,-2)\over 2}, \qquad Y_{10} = J(3,-3),  \qquad Y_{11}= 2 J(3,-4).    \nonumber
\eea

The respective codes can be found on the web page of the ITEP Lattice group \\
{ \tt http://www.lattice.itep.ru/$\sim$pbaivid/lattpt/}. \ 
The results stored there are as follows: 
(i) the program for a computation of $F(p,q;\tilde  n)$ at $0\leq p,q \leq 9$ and $n_1+n_2+n_3+n_4 \leq 25$;
(ii) the values of the functions $J(p,q)$ and $B(p,q)$ at $-26 \leq p \leq 0,\ \ -56 - 2p \leq q \leq 34 $ 
and the values of $J(p,q)$  at $1\leq p \leq 9, \ \ -28 \leq q \leq 33 - p$; and
(iii) The explicit expressions for $F(p,q;\tilde n)$
at some particular values of $p$ and $q$ and all $n_1\leq 6$.

\section{The L{\" u}scher--Weisz method}

\noindent To outline the LW method \cite{Luscher}
of computation of two-loop diagrams in the coordinate representation, 
we consider the diagram in~Fig.1, given by the expression

\hspace*{-2mm}\begin{minipage}{0.4\hsize}
\begin{picture}(170,130)(0,0)
\Text(-6,96)[tr]{\large $x$} 
\Text(172,96)[tr]{\large $0$} 
\ArrowArcn(80,0)(113,135,45)
\ArrowArc(80,160)(113,225,315)
\ArrowLine(0,80)(160,80)
\end{picture} 
\vspace*{-9mm}
\bc
Figure 1
\ec
\end{minipage}\hfill\begin{minipage}{0.52\hsize}
\be\label{DiagSunSet00}
A_B(p) = \sum_{x\in\Lambda} e^{-ipx} G_B^3(x).
\ee
In the bosonic case, L{\" u}scher and Weisz
calculated $A_B(0)$ and its asymptotic expansion when $p \to 0$;
they used the following representation:
\bea\label{DivLambdaIn2Domains}
A_B(0) &=& G_B^3(0) + \sum_{x\in \Lambda'} G_{as}^3(x) \\ \nn
&& +  \sum_{x\in\{\cF_{N}\}} \Big( G_B^3(x) - G_{as}^3(x)\Big) \\ \nn
&& +  \sum_{ x\in\{\Lambda'  \backslash \cF_{N}\}} \Big( G_B^3(x) - G_{as}^3(x)\Big), \nn
\eea
\end{minipage}

\noindent where           
$\cF_{N} =\{ x: |x_1| + |x_2| + |x_3| + |x_4| \leq N \} $,
and $G_{as}(x)$ is an asymptotic approximation of $\ds G_B(x)$ when $x\to \infty$, 
\be\label{DBas}
   G_{as}(x)  = {1\over [x^2]  }
	\; + \; \left(  {2 [x^4] -  [x^2]^2 \over [x^2]^4   }  \right)
  + \left( 40 \; {[x^4]^2 \over [x^2]^7 }
          + 16 \; {[x^4] \over [x^2]^5 }
          - 48 \; {[x^6] \over [x^2]^6 }
          - 4\; {1 \over [x^2]^3 }           \right) + ... 
\ee

In the domain $\cF_{N}$, the propagator $\ds G_B(x)$ can computed by the recursion formulas
\be\label{RecRelxBosProp}
G_B(x+\hat \mu) = G_B(x - \hat \mu) + \frac{2x_\mu}{\Big(\sum_{\nu=1}^4 x_\nu\Big)}\; \sum_{\lambda=1}^4 (G_B(x)- G_B(x-\hat \lambda)),
\ee
which allow to express it in terms of $G_B(0,0,0,0)=Z_0$ and $G_B(1,1,0,0) = - 1/4 + Z_1 + Z_0$.
The domain $\{\Lambda \backslash \cF_{N} \}$ is chosen so that the propagator 
is fitted by its asymptotic expression (\ref{DBas}) with a sufficient precision 
making it possible to neglect the third sum in the formula (\ref{DivLambdaIn2Domains}).
Then the first sum can be calculated exactly using the summation formulas derived in \cite{Luscher}
and the second sum can be expressed in terms of $Z_1$ and $Z_2$ by employing 
the relations (\ref{RecRelxBosProp}).

It should be emphasized that $A_B(0)$ is the coefficient of the expansion
$$
A_B(p) = {1\over a^2}\Big( A_B(0) + (pa)^2 \left[ A_1 + B_1 \ln(pa)^2 \right] + O((pa)^4) \Big),
$$
where $a$ is the length of a link; that is, $A_B(0)$ is the coefficient of the 
divergent part. It does not vanish, though the ``bigmac'' diagram 
depicted in Fig.1 converges in the dimensional regularization
provided that all masses and the external momentum are equal to zero.

\section{Two-loop fermionic integrals.}

In the fermionic case, calculations are performed by the same procedure, however,
{\bf we have no recursion relations similar to (\ref{RecRelxBosProp}).}
The fermionic propagator in $x$-representation 
\be 
G_F(x_1,x_2,x_3,x_4)=
\int {d^4k\over (2\pi)^4} 
{\cos(k_1 x_1) \cos(k_2 x_2) \cos(k_3 x_3) \cos(k_4 x_4) \over 
\Delta_F}
\ee
is expressed in terms of the quantities
\be 
F(p,q;n_1,n_2,n_3,n_4)=
\int {d^4k\over (2\pi)^4} 
{\cos^{n_1}(k_1) \cos^{n_2}(k_2) \cos^{n_3}(k_3) \cos^{n_4}(k_4) \over \Delta_B^q
\Delta_F^{p+\delta}}
\ee 
by making use of the relations
\be 
\cos (nx) = 2^{n-1} \cos^n x \;+\; {n\over 2} \sum_{k=0}^{[n/2]-1}
{(-1)^{k+1} \over k+1} C_{n-k-2}^{k} (2\cos x)^{n-2k-2}.
\ee
To employ the LW method outlined above, we compile a table of values of $G_F(x)$ over the domain 
$ x_1\geq x_2\geq x_3\geq x_4\geq 0,\  |x| \leq 48$
and derive an asymptotic approximation of $G_F(x)$ at $|x| \to \infty$ up to the
terms of the order $\ds 1/[x^2]^4$.
To treat integrals with nontrivial numerators, we should also compile the tables of the values
\be \label{GKL-definitions}
K_{B[F]} = \int {dp \over (2\pi^4)} {(e^{-ipx}\; - 1) \over D_{B[F]}^2 (p) } \ , \quad
L_{B[F]} = \int {dp \over (2\pi^4)} {\ds \left( e^{-ipx} - 1 + {x^2\over 8} \; (4\; -\; \sum_{\mu=1}^4 \cos^2 k_\mu )\;\right)\over D_{B[F]}^3 (p) } \ , 
\ee
Each of these tables involves 14147 entries, each entry is a linear combination of the  constants
$F_0, Z_0, Z_1$, $Y_0, Y_1, ... Y_{11}$, $\ds {1\over(2\pi)^2}$, and 1 with
rational coefficients; from 5 to 20~MB per table in size.
Fortunately, they can be conveniently treated with FORM \cite{FORM}.

The precision of 20 significant digits in determination of the constants 
$Y_4\div Y_{11}$ \cite{CapitaniPR}, \cite{Burgio}
is not sufficient for computation of $G_F(x)$ at $|x| > 6$. 
Using the procedure proposed in \cite{Luscher} for calculation of $Z_0$ and $Z_1$,
we obtain
\bea\nn
Y_4    &=&  0.08539036359532067913516702888533412058194147127443265(1) \\ \nonumber
Y_5    &=&  0.46936331002699614475347539705751803482046295887523184(1) \\ \nonumber
Y_6    &=&  3.39456907367713000586008689702374496453685272313733503(1) \\ \nonumber
Y_7    &=&  0.05188019503901136636490228766471579940968012757291508(1) \\ \nonumber
Y_8    &=&  0.23874773756341478520233613930386970445280194983477988(1) \\ \nonumber
Y_9    &=&  0.03447644143803223145396188144243193600121277124715784(1) \\ \nonumber
Y_{10} &=&  0.13202727122781293085314731098196596971197144795959477(1) \\ \nonumber
Y_{11} &=&  0.75167199030295682253543148590778110991011277193144803(1)  \nonumber
\eea 

At $|x|>48$, $G_F(x)$ is approximated by the function
\be   G^{(as)}_F(x) = {1\over [x^2] }
	  + \left( {8  [x^4] - 4  [x^2]^2  \over [x^2]^4 }   \right)
  + \left(  640 \; {[x^4]^2 \over [x^2]^7 }  - 768 \; {[x^6] \over [x^2]^6 }
          + 208 \; {[x^4] \over [x^2]^5 }
          -  {40 \over [x^2]^3 }          \right) + ...,
\ee
To provide an example, let us consider the following two-loop fermionic integrals:
\bea
Q_1^{BBB} &=& \int_{BZ} {d^4k \over (2\pi)^4} \; {d^4q \over (2\pi)^4} \sum_{\mu=1}^4 
{\hat k_\mu^2 \hat q_\mu^2 \over D_B(k) D_B(q) D_B(r) } \\ \nonumber
Q_1^{BBF} &=& \int_{BZ} {d^4k \over (2\pi)^4} \; {d^4q \over (2\pi)^4} \sum_{\mu=1}^4 
{{\hat k_\mu^2} {\hat q_\mu^2} \over D_B(k) D_B(q) D_F(r) } \\ \nonumber
\eea
and similar quantities with other combinations of bosonic and fermionic propagators. 
We can also consider
\be
Q_2^{BBF} = \int_{BZ} {d^4k \over (2\pi)^4} \; {d^4q \over (2\pi)^4} \sum_{\mu=1}^4 
{{\hat k_\mu^2} {\hat q_\mu^2} {\hat r_\mu^2} \over D_B(k) D_B(q) D_F(r) } 
\ee
etc. The results of the computations are as follows:
\bea
&Q_1^{BBB}  = 0.042306368(1) \qquad \qquad   &Q_1^{FBB}  = 0.020079702(3) \\ \nonumber
&Q_1^{BBF}  = 0.024555253(3) \qquad \qquad   &Q_1^{FFB}  = 0.00969896(1)  \\ \nonumber
&Q_1^{BFF}  = 0.01173224(1)  \qquad \qquad   &Q_1^{FFF}  = 0.00576013(3)  \\ \nonumber
&Q_2^{BBB}  = 0.05462397818(1) \qquad \qquad &Q_2^{BBF}  = 0.02659175158(3) \\ \nonumber
&Q_2^{BFF}  = 0.0130373237(1)  \qquad \qquad &Q_2^{FFF}  = 0.0064945681(3)   \nonumber
\eea

\section{Summary and Outlook}

The BCP algorithm has been realized on a computer.
The basic fermionic integrals $G(p,q)$ are found over a sufficiently large domain of values of $p$ and $q$.
This allows (i) to express one-loop intergals involving fermionic denominators in terms of
the constants $F_0, Z_0, Z_1$ and $Y_0\div Y_{11}$ and (ii) to compute $G_F(x)$ at $|x|\leq 96$.

The LW method is extended to the case of fermions;
asymptotic behavior of the fermionic propagator at $|x|\to \infty$ is found.
Therewith, $G_F(x)$ is expressed at $|x|\leq 48$ in terms of the constants $Y_4\div Y_{11}$,
the values of which are computed to a precision of 54 significant digits.
This is really needed for calculation of two-loop integrals.
A new feature of FORM - a possibility to work with database-like structures -
proved to be useful for summation over the domain $|x|\leq 48$.
As an illustration, several two-loop fermionic integrals are evaluated 
at zero external momentum.

Operations with a table of precise values of the functions $G_{B(F)}(x)$
$K_{B(F)}(x)$ and $L_{B(F)}(x)$ allow to compute one-loop and two-loop 
diagrams of the propagator type at nonvanishing external momentum.
The work is in progress!
\vspace*{3mm}

{\bf Acknowledgments:} I am grateful to V.Bornyakov, A.Kataev, H.Perlt, and A.Schiller
for stimulating discussions. This work was supported in part by 
the Russian Foundation for Basic Research (grant no. 07-02-00237-a)
and by the grant for scientific schools NSh-6260.2010.2.

\end{document}